\documentclass[11pt]{article}
\usepackage{amssymb}
\usepackage{graphicx}
\usepackage{lscape}
\usepackage{theorem}

\newcommand \be  {\begin{equation}}
\newcommand \bea {\begin{eqnarray}}
\newcommand \ee  {\end{equation}}
\newcommand \eea {\end{eqnarray}}
\newcommand \E {{\rm E}}

\renewcommand \epsilon {\varepsilon}

\theoremheaderfont{\scshape}
\theoremstyle{break}
\theorembodyfont{\itshape}
\newtheorem{theorem}{Theorem}
\newtheorem{lemma}{Lemma}

\newtheorem{corollary}{Corollary}

\begin{document}

\title{Tail Dependence of Factor Models
\footnote{We acknowledge helpful discussions and
exchanges with J.P. Laurent and V. Pisarenko. This work was partially
supported by
the James S. Mc Donnell Foundation 21st century scientist award/studying
complex system.}}
\author{Y. Malevergne$^{1,2}$ and D. Sornette$^{1,3}$ \\
\\
$^1$ Laboratoire de Physique de la Mati\`ere Condens\'ee CNRS UMR 6622\\
Universit\'e de Nice-Sophia Antipolis, 06108 Nice Cedex 2, France\\
$^2$ Institut de Science Financi\`ere et d'Assurances - Universit\'e Lyon I\\
43, Bd du 11 Novembre 1918, 69622 Villeurbanne Cedex, France\\
$^3$ Institute of Geophysics and Planetary Physics
and Department of Earth and Space Science\\
University of California, Los Angeles, California 90095, USA\\
\\
email: Yannick.Malevergne@unice.fr and sornette@unice.fr\\
fax: (33) 4 92 07 67 54\\
}

\date{03/02/2002}

\maketitle

\begin{abstract}
Using the framework of factor models, we establish the general expression
of the coefficient
of tail dependence between the market and a stock (i.e., the probability that
the stock incurs a large loss, assuming that the market has also undergone a
large loss) as a function of the parameters of the underlying factor model and
of the tail parameters of the distributions of the factor and of the
idiosyncratic
noise of each stock. Our formula
holds for arbitrary marginal distributions and in addition
does not require any parameterization of the multivariate
distributions of the market and stocks. The
determination of the extreme parameter, which is not accessible by a direct
statistical inference, is made possible by the measurement of parameters whose
estimation involves a significant part of the data with sufficient statistics.
Our empirical tests find a good agreement between the calibration of the tail
dependence coefficient and the realized large losses over the period from 1962
to 2000. Nevertheless, a bias is detected which suggests the presence of an
outlier in the form of the crash of October 1987. \end{abstract}


\section*{Introduction}

The concept of extreme or ``tail dependence'' probes the reaction of
a variable to the
realization of another variable when this realization is of extreme
amplitude and very low
probability.
The dependence, and especially the extreme dependence, between two
assets or between an asset and any other exogeneous economic variable is an
issue of major importance both for practioners and for academics.
The determination of extreme dependences is
crucial for financial and for insurance institutions involved in
risk management. It is also fundamental
for the establishment of a rational investment policy striving for the best
diversification of the various sources of risk. In all these situations, the
objective is to prevent or at least minimize
the simultaneous occurrence of large losses
across the different positions held in the portfolio.

  From an academic perspective, taking into account
the extreme dependence properties provide
useful yardsticks and important constraints for the construction of models,
which should not underestimate or overestimate risks. From the point of view
of univariate statistics, extreme values theory provides the mathematical
framework for the classification and quantification of very large risks.
It has been used for instance to study the
phenomenon of contagion (as an example, see \cite{LS01} for a study
of contagion
across international equity markets). This has been made possible
by the existence of a ``universal'' behavior summarized by the
Gnedenko-Pickands-Balkema-de Haan theorem which gives a natural limit
law for peak-over-threshold values in the form of the Generalized
Pareto Distribution
\cite[pp 152-168]{Embrechtsbook}. In constrast, no
such result is yet available in the multivariate case. In such absence
of theoretical guidelines, the alternative is therefore to impose
some dependence structure in a rather ad hoc and arbitrary way. This was
the stance taken for instance in \cite{LS01}.

This approach, where the
dependence structure is not determined from empirical facts or from
an economic model, is not fully satisfying.
As a remedy, we propose a new approach, which does not directly rely on
multivariate extreme values theory, but rather derives the extreme
dependence structure from the characteristics of a financial model of assets.
Specifically, we use the general class of factor models, which is probably one
of the most versatile and relevant one, and
whose introduction in finance can be traced back at least to
\cite{R76}. The factor models are now widely used in many branches of finance,
including stock return models, interest rate models \cite{V77, BS78, Cox_etal},
credit risks models \cite{C98, G00, Lucas_etal}, etc., and are
found at the core of many theories and equilibrium models.

Here, we will focus our efforts on the characterization
of the extreme dependence between stock returns and
the market return. The role of the market return as a factor explaining the
evolution of individual stock returns is supported both by theoretical models
such as the Capital Asset Pricing Model \cite{S64, L65, M66} or the
Arbitrage Pricing Theory \cite{R76} and by empirical studies \cite{FMB73, KS87}
among many others. It has even been shown in \cite{R88} that in certain
dramatic circumstances, such as the October 1987 stock-market crash, the
(global) market was the sole relevant factor needed to explain the stock market
movements and the propagation of the crash across countries.
Thus, the choice of factor models is a very natural starting point for
studying extreme dependences from a general point of view. The main
gain is that, without imposing any {\it a priori} ad hoc dependences other than
the definition of the factor model, we shall be able to
derive the general properties of extreme dependence between an asset
and one of its factor
and to empirically determine these properties by a simple estimation
of the factor
model parameters.

The plan of our presentation is as follows.
Section 1 defines the concepts needed for the characterization
and quantification of extreme dependences. In particular, we recall
the definition of
the coefficient of tail dependence, which captures in a single number the
properties of extreme dependence between two random variables: the
tail dependence is
defined as the
probability for a given random variable to be large assuming that another
random variable is large, at the same probability level. We shall also need
some basic notions on
dependences between random variables using the mathematical
concept of copulas. In order to provide some perspective on the
following results,
this section also contains the expression of some classical exemples
of tail dependence
coefficients for specific multivariate distributions.

Section 2 states our main result in the form of a general theorem allowing the
calculation of the coefficient of tail dependence for any factor model with
arbitrary distribution functions of the factors and of the idiosyncratic noise.
We find that the factor must have sufficiently ``wild'' fluctuations (to be
made precise below) in
order for the tail dependence not to vanish. For normal distributions of
the factor, the tail
dependence is identically zero, while for regularly varying distributions
(power laws), the tail dependence is in general non-zero.

Section 3 is devoted to the empirical estimation of the
coefficients of tail dependence between individual stock returns and the
market return. The tests are performed for daily stock returns. The estimated
coefficients of tail dependence are found in good agreement with the fraction
of historically realized extreme events that occur simultaneously with any of
the ten largest losses of the market factor (these ten largest losses were not
used to calibrate the tail dependence coefficient). We also find some
evidence for comonotonicity in the crash of Oct. 1987, suggesting that
this event is an ``outlier,'' providing additional support to a
previous analysis of
large and extreme drawdowns.

We summarize our results and conclude in section 4.


\section{Intrinsic measure of casual and of extreme dependences}

This section provides a brief informal summary of the mathematical concepts
used in this paper to characterize the normal and extreme dependences
between asset returns.

\subsection{How to characterize uniquely the full dependence between two
random variables?}

The answer to this question is provided by the mathematical notion
of ``copulas,'' initially introduced by
\cite{S59} \footnote{The reader is refered to \cite{J97, FV98} or
\cite{Nelsen} for a detailed survey of the notion of copulas and a
mathematically rigorous description of their properties.}, which allows one to
study the dependence of random
variables independently of the behavior of their marginal distributions.
Our presentation focuses on two variables
but is easily extended to the case of $N$ random variables, whatever
$N$ may be.
Sklar's Theorem states that, given the joint distribution function
$F(\cdot, \cdot)$
of two random variables $X$ and $Y$ with marginal
distribution $F_X(\cdot)$ and $F_Y(\cdot)$ respectively, there exists a
function $C(\cdot, \cdot)$ with range in $[0,1] \times [0,1]$ such that
\be
F(x,y) = C(F_X(x),F_Y(y))~,
\ee
for all $(x,y)$. This function $C$ is the {\it copula} of the two random
variables $X$ and $Y$, and is unique if the random variables have continous
marginal distributions.
Moreover, the following result shows that
copulas are intrinsic measures of dependence. If
$g_1(X), g_2(Y)$ are strictly increasing on the ranges of
$X, Y$, the random variables $\tilde X=g_1(X),
\tilde Y=g_2(Y)$ have exactly the same copula $C$ \cite{Lindskog}.  The
copula is thus invariant under strictly increasing transformation of
the variables.  This provides a powerful way of studying
scale-invariant measures of associations.  It is also a natural
starting point for construction of multivariate distributions.

\subsection{Tail dependence between two random variables}
\label{sec:ctd}

A standard measure of dependence between two random variables is
provided by the
correlation coefficient. However, it suffers from at least three deficiencies.
First, as stressed by \cite{EMN99}, the correlation coefficient is an
adequate measure of dependence only for
elliptical distributions and for events of moderate sizes.
Second, the correlation coefficient measures only the degree of linear
dependence and does not account of any other nonlinear functional
dependence between the
random variables. Third, it agregates both the marginal behavior of each
random variable and their dependence. For instance, a simple change
in the marginals implies in general  a
change in the correlation coefficient, while the copula and, therefore the
dependence, remains unchanged.
Mathematically speaking, the correlation
coefficient is said to lack the property of invariance under
increasing changes of variables.

Since the copula is the unique and intrinsic measure of dependence,
it is desirable to define measures of dependences which depend only on the
copula. Such measures have in fact been known for a long
time. Examples are provided by the concordance measures, among which
the most famous are
the Kendall's tau and the Spearman's rho (see \cite{Nelsen} for a detailed
exposition). In particular, the Spearman's rho quantifies the
degres of functional dependence between two random variables: it equals one
(minus one) when and only when the first variable is an increasing
(decreasing) function of the second variable. However, as for the correlation
coefficient, these concordance measures do not provide a useful measure of the
dependence for extreme events, since they are constructed over the
whole distributions.

Another natural idea, widely used in the contagion literature, is to
work with the
conditional correlation coefficient, conditioned only on the largest events.
But, as stressed by \cite{Boyer_etal}, such conditional correlation
coefficient suffers from a bias: even for a constant {\it unconditional}
correlation coefficient, the {\it conditional} correlation coefficient changes
with the conditioning set.
Therefore, changes in the conditional correlation do not provide a
characteristic
signature of a change in the true correlations. The conditional concordance
measures suffer from the same problem.

In view of these deficiencies, it is natural to come back to a
fundamental definition of dependence through the use of probabilities.
We thus study the conditional probality that the first variable is
large conditioned on the
second variable being large too: $\bar F(x|y)=\Pr\{X>x | Y >y\}$, when $x$ and
$y$ goes to infinity. Since the convergence of $\bar F(x|y)$ may depend on the
manner with which $x$ and $y$ go to infinity (the convergence is not
uniform), we need to
specify the path taken by the variables to reach the infinity.
Recalling that it
would be preferable to have a measure
which is independent of the marginal distributions of $X$ and $Y$,
it is natural to reason in the quantile space. This leads to choose
$x={F_X}^{-1}(u)$ and $y={F_Y}^{-1}(u)$ and replace the conditions
$x, y \to \infty$ by $u \to 1$. Doing so, we define
the so-called coefficient of upper tail dependence
\cite{Coles_etal,Lindskog,EMS01}:
\be
\lambda_+ = \lim_{u \rightarrow 1^-}  \Pr\{X> {F_X}^{-1}(u)~|~ Y
>{F_Y}^{-1}(u)\}~.
\label{gjjtr}
\ee
As required, this measure of dependence is independent of the marginals, since
it can be expressed in term of the copula of $X$ and $Y$ as
\be
\lambda_+ = \lim_{u \rightarrow 1^-} \frac{1-2u+C(u,u)}{1-u}~.
\ee
This representation shows that $\lambda_+$ is symmetric in $X$ and
$Y$, as it should for a reasonable measure of dependence.

In a similar way, we define the coefficient of lower tail dependence as
the probabilty that $X$ incurs a large loss
assuming that $Y$ incurs a large loss at the same probability level
\be
\lambda_- = \lim_{u \rightarrow 0^+}  \Pr\{X< {F_X}^{-1}(u)~|~ Y
<{F_Y}^{-1}(u)\} = \lim_{u \rightarrow 0^+} \frac{C(u,u)}{u}~.
\ee

The values of the coefficients of tail dependence are known explicitely for
a large number of different
copulas. For instance, the Gaussian copula, which is the copula derived from de
Gaussian multivariate distribution, has a zero coefficient of tail dependence.
In contrast, the Gumbel's copula used by \cite{LS01} in the study of the
contagion between international equity markets, which is defined by
\be
C_\theta(u,v) = \exp \left( - \left[(-\ln u ) ^\theta + (- \ln v)^\theta
\right]^\frac{1}{\theta} \right), ~~~~~~~~~ \theta \in[0,1],
\ee
has an upper tail coefficient
$\lambda_+=2-2^\theta$. For all $\theta$'s smaller than one, $\lambda_+$ is
positive and the Gumbel's copula is said to present tail dependence, while for
$\theta=1$, the Gumbel copula is said to be asymptotically independent. One
should however use this terminology with a grain of salt as ``tail
independence''
(quantified by $\lambda_+=0$ or $\lambda_-=0$) does not imply necessarily that
large events occur independently (see \cite{Coles_etal} for a precise
discussion of
this point).


\section{Tail dependence of factor models}
\label{sec:th}

\subsection{General result}

We now state our main theoretical result.
Let us consider two random variables $X$ and $Y$ of cumulative
distribution functions $F_X(X)$ and $F_Y(Y)$, where $X$ represents
the return of a single stock and $Y$ is the market return. Let us
also introduce an idiosyncratic noise $\epsilon$, which is
assumed independent of the market return $Y$. The factor model is defined by
the following relationship between the individual stock return $X$, the market
return $Y$ and the idiosyncratic noise $\epsilon$:
\be
\label{eq:FM}
X = \beta \cdot Y + \epsilon~.
\ee
$\beta$ is the usual coefficient introduced by the Capital Asset
Pricing Model \cite{S64}. Let us stress that $\epsilon$ may embody other
factors $Y'$, $Y'',\ldots$, as long as they remain independent of
$Y$. Under such conditions and a few other technical assumptions detailed
in the theorem established in appendix \ref{app:1}, the  coefficient of (upper)
tail dependence between $X$ and $Y$ defined in (\ref{gjjtr}) is obtained as
\be
\lambda_+ = \int_{\max \left\{ 1, \frac{l}{\beta} \right\} }^\infty dx ~f(x)~,
\label{fundtheorem}
\ee
where $l$ denotes the limit, when $u \to 1$, of the ratio
${F_X}^{-1}(u)/{F_Y}^{-1}(u)$, and $f(x)$ is the limit, when $t \to +\infty$,
of $t \cdot P_Y(t x) / \bar F_Y(t)$. $P_Y$ is the distribution density of $Y$
and $\bar F_Y =1-F_Y$ is the complementary cumulative distribution
function of $Y$.
A similar expression obviously holds, {\it mutatis mutandis}, for the
coefficient of lower tail dependence.

We now derive two direct consequences of this result (\ref{fundtheorem})
(see corollary 1 and 2 in appendix \ref{app:1}), concerning rapidly
varying and regularly varying factors.

\subsection{Absence of tail dependence for rapidly varying factors}

Let us assume that the factor $Y$ and
the idiosyncratic noise $\epsilon$ are normally distributed
(the second assumption is made for simplicity and will be relaxed below).
As a consequence, the joint distribution of $(X,Y)$
is the bivariate Gaussian distribution. Refering to the results stated in
section \ref{sec:ctd}, we conclude that the copula of $(X,Y$) is the
Gaussian copula whose coefficient of tail dependence is zero. In fact, it is
easy to show that $\lambda=0$ for any distribution of $\epsilon$.

More generally, let us assume that the distribution of the factor
$Y$ is rapidly varying, which describes the Gaussian, exponential and any
distribution decaying faster than any power-law. Then,
the coefficient of tail dependence is identically zero.
This result holds for any arbitrary distribution
of the idiosyncratic noise (see corollary 1 in apendix \ref{app:c1}).

This statement is somewhat counter-intuitive since one could expect
{\it a priori} that the
coefficient of tail dependence does not vanish as soon as the tail of the
distribution of factor returns is fatter than the tail the distribution
noise returns. However,
this example indicates that this is not the case and, in order to get
a non-vanishing
tail-dependence, the fluctuations
of the factor must be 'wild' enough, which is not realized with rapidly varying
distributions.

\subsection{Coefficient of tail dependence for regularly varying factors}
\subsubsection{Example of the factor model with Student distribution}

In order to account for the power-law tail behavior observed for the
distributions of assets returns it is logical to consider that the factor and
the indiosyncratic noise also have power-law tailed distributions. As an
illustration, we will assume that $Y$ and $\epsilon$ are distributed according
to a Student's distribution with the same number of degrees of freedom $\nu$
(and thus same tail exponent $\nu$).
Let  us denote by $\sigma$ the scale factor of the distribution of $\epsilon$
while the scale factor of the distribution of $Y$ is choosen equal to
one\footnote{Such a choice is always possible via a rescaling of the
coefficient $\beta$.}. Applying the theorem previously established, we find
that $f(x)=\nu/x^{\nu+1}$ and $l=  \beta ~ \left[ 1+ \left(
\frac{\sigma}{\beta} \right)^\nu \right]^{1/\nu}$, so that the coefficient of
tail dependence is
\be
\label{eq:8}
\lambda_\pm = \frac{1}{1+\left( \frac{\sigma}{\beta}
\right)^\nu},~~~~~~~~\mbox{and}~~\beta>0.
\ee
As expected, the
tail dependence increases as $\beta$ increases and as $\sigma$ decreases.
The dependence with respect to $\nu$ is less intuitive. In particular, let
$\nu$ go to infinity. Then, $\lambda \to 0$ if $\sigma >
\beta$  and $\lambda \to 1$ for $\sigma < \beta$. This is surprising as
one could argue that, as $\nu \to \infty$, the Student distribution
tends to the Gaussian law. As a consequence, one would expect
the same coefficient of
dependence $\lambda_\pm = 0$ as for rapidly varying functions.
The reason for the non-certain convergence of $\lambda_\pm$ to zero as
$\nu \to \infty$ is rooted in a subtle non-commutativity
(and non-uniform convergence) of the two limits
$\nu \to \infty$ and $u \to 1$. Indeed, when taking first the limit
$u \to 1$, the result
$\lambda \to 1$ for $\beta > \sigma$ indicates that a sufficiently strong
factor coefficient $\beta$ always ensures the validity of
the power law regime, whatever the value of $\nu$. Correlatively, in
this regime
$\beta > \sigma$, $\lambda_\pm$ is an increasing function of $\nu$.

\subsection{General result}

We now provide the general result valid for any regularly
varying distribution. Let the factor $Y$ follows a regularly varying
distribution with tail index $\alpha$: in other words, the complementary
cumulative distribution of $Y$ is such that $\bar F_Y(y) = L(y) \cdot
y^{-\alpha}$, where
$L(y) $ is a slowly varying function, i.e:
\be
\lim_{t \rightarrow \infty} \frac{L(ty)}{L(t)} = 1,~~~~~~\forall y>0.
\ee
Corollary 2 in appendix \ref{app:c2} shows that
\be
\lambda = \frac{1}{\left[ \max \left\{ 1, \frac{l}{\beta}
\right\}\right]^\alpha}~,
\ee
where $l$ denotes the limit, when $u \to 1$, of the ratio
${F_X}^{-1}(u)/{F_Y}^{-1}(u)$.
In the case of particular interest when the distribution of $\epsilon$ is also
regularly varying with tail index $\alpha$ and if, in addition, we have $\bar
F_Y(y) \sim {C_y} \cdot y^{-\alpha}$ and  $\bar F_\epsilon(\epsilon) \sim
{C_ \epsilon} \cdot \epsilon^{-\alpha}$, for large $y$ and $\epsilon$, then
the coefficient of tail dependence is a simple function of the ratio
$C_\epsilon / C_y$ of the scale factors:
\be
\label{eq:ctd}
\lambda = \frac{1}{1+ \beta^{-\alpha} \cdot \frac{C_\epsilon}{ C_y}}~.
\ee

When the tail indexes $\alpha_Y$ and $\alpha_{\epsilon}$
of the distribution of the factor and the residue are
different, then $\lambda=0$ for $\alpha_Y < \alpha_{\epsilon}$
and $\lambda=1$ for $\alpha_Y > \alpha_{\epsilon}$.

Now that we have entirely characterized the tail dependence for the factor
model, we will use these results to estimate
empirically the tail dependence between
different stock returns and the market return and test our prediction
on historical events.


\section{Empirical study}

We now apply our theoretical results to the daily returns of a
set of stocks traded on the New York Stock Exchange. In order to estimate
the parameters of the factor model (\ref{eq:FM}), the Standard and
Poor's $500$ index, which represents about 80\% of the total market
capitalization, is
choosen to represent the common ``market factor.''

We describe the set of selected stocks in the next sub-section. Next, we
estimate the parameter $\beta$ in (\ref{eq:FM}) and check the independence of
the market returns and the residues. Then, applying the commonly used
hypothesis according to which the tail of the distribution of assets return is
a power law (see \cite{L96, Lux96,P96,Gopikrishnan}), we estimate the tail
index and the scale factor of these distributions, which allows us to calculate
the coefficients of tail dependence between each asset return and the market
return. Finally, we perform an analysis of the historical data to check the
compatibility of our prediction on the fraction of realized large losses of the
assets that occur simultaneously with the large losses of the market.

The results of our analysis are reported below in terms
of the returns rather than in terms of the excess returns above the
risk free interest rate, in apparent contradiction with the prescription of the
CAPM. However, for daily returns, the difference between returns and excess
returns is negligible. Indeed, we checked that neglecting the
difference between the returns and the excess returns does not affect
our results
by re-running all the study described below in terms of the
excess returns and found that the tail dependence did not change by more than
$0.1\%$.

\subsection{Description of the data}

We study a set of twenty assets traded on the New York Stock
Exchange. The criteria presiding over the selection of the assets
(see column 1 of table \ref{table:asset}) are that (1) they
are among the stocks with the largest capitalizations, but (2)
each of them should have a weight smaller than $1\%$ in the Standard
and Poor's $500$ index,
so that the dependence studied here
does not stem trivially from their overlap with the market factor (taken as
the Standard and Poor's $500$ index).

The time interval we have considered ranges from July 03,
1962 to December 29, 2000, corresponding to 9694 data points, and represents
the largest set of daily data available from the Center for Research in
Security Prices (CRSP) . This large time interval is important to let
us collect as many
large fluctuations of the returns as is possible in order
to sample the extreme tail dependence.
Moreover, in order to allow for a non-stationarity over the four
decades of the study,
to check the stability of our results and to test
the stationnarity of the tail dependence over the time, we
split this set into two subsets. The first one ranges from July
1962 to December 1979, a period with few very large return
amplitudes, while the second
one ranges from January 1980 to December 2000, a period which witnessed several
very large price changes (see
table \ref{table:asset}) which shows the good stability of the
standard deviation between
the two sub-periods while the higher
cumulants such as the excess kurtosis often increased dramatically in
the second sub-period
for most assets).
The table \ref{table:asset} presents the main statistical properties of our set
of stocks during the three time intervals. All
assets exhibit an excess kurtosis significantly different from zero
over the three
time interval, which is inconsistent with the assumption of Gaussianly
distributed returns. While the standard deviations remain stable
over time, the excess kurtosis increases significantly from the first
to the second period. This is in resonance with
the financial community's belief that stock price
volatility has increased over time, a still controversial result 
\cite{Joneswilson}.

\subsection{Calibration of the factor model}

The determination of the parameters
$\beta$ and of the residues $\epsilon$ entering in the definition of
the factor model (\ref{eq:FM}) is performed for each asset by
regressing the stocks returns on the market return. The coefficient
$\beta$ is thus given by the ordinary least square estimator, which
is consistent as long as the residues are white noise and with zero
mean and finite variance. The
idiosyncratic noise $\epsilon$ is obtained by substracting $\beta$
times the market return to the stock return. Table \ref{table:param}
presents the results for the three periods we consider. For each
period, we give
the value of the estimated coefficient $\beta$ and the correlation coefficient
between the market returns and the idiosyncratic noise.  A Fisher's test is
shows that, at the $95\%$ confidence level, none af these
correlation coefficients is significantly different from zero. This does not
necessarily ensures the independence of the idiosyncratic noises with
respect to the
market return, but is nonetheless a positive result for the validity of the
factor decomposition (\ref{eq:FM}).

The coefficient $\beta$'s we obtain by regressing
each asset returns on the Standard \& Poor's 500 returns are very
close to within
their uncertainties to the $\beta$'s given by the CRSP database, which are
estimated by regressing the assets returns on the value-weighted market
portfolio. Thus, the choice of the Standard and Poor's 500 index to represent
the whole market portfolio is reasonable.

\subsection{Estimation of the tail indexes}

Assuming that the distributions of stocks and market returns are
asymptotically power laws  \cite{L96, Lux96,P96,Gopikrishnan},
we now estimate the tail index
of the distribution of each stock and their corresponding residue
by the factor model, both for the positive and negative
tails. Each tail index $\alpha$ is given by Hill's etimator:
\be
\hat \alpha = \left[ \frac{1}{k} \sum_{j=1}^k \log x_{j,N} - \log
    x_{k,N} \right]^{-1},
\ee
where $x_{1,N} \ge x_{2,N} \ge \cdots \ge x_{N,N}$ denotes the ordered
statistics of the sample containing $N$ independent and identically
distributed realizations of the variable $X$.

Hill's estimator is asymptotically normally distributed with mean $\alpha$ and
variance $\alpha^2/k$. But, for finite $k$, it is known that the
estimator is biased. As
the range $k$ increases, the variance of the estimator decreases while its bias
increases. The competition between these two effects implies that there is an
optimal choice for $k=k^*$ which minimizes the mean squared error of
the estimator. To select this value $k^*$, one can apply the \cite{DdV97}'s
algorithm which is an improvement over the \cite{H90}'s subsample bootstrap
procedure. One can also prefer the more recent \cite{DdHPdV2000}'s algorithm
for the sake of parsimony. We have tested all three algorithms to
determine the optimal $k^*$. It turns out that the \cite{DdHPdV2000}'s
algorithm developped for high frequency data is not well adapted to samples
containing less than 100,000 data points, as is the case here. Thus, we have
focused on the two other algorithms. An accurate determination of $k^*$ is
rather difficult with any of them, but in every case, we found that the
relevant range for the tail index estimation was between the $1\%$ and
$5\%$ quantiles. Tables \ref{table:tail2} and \ref{table:tail3} give
the estimated
tail index for each asset and residues at the $1\%$, $2.5\%$ and
$5\%$ quantile, for both the positive and the negative tails for the
two time sub-intervals.
The second time interval from January 1980 to December 2000 is characterized
by values of the tail indexes that are
homogeneous over the various quantiles and range between 3 and 4 for the
negative tails and between 3 and 5 for the positive tails. There is
slightly more
dispersions in the first time interval from July 1962 to December 1979.

For each asset and their residue of the regression on the market factor,
we tested whether the hypothesis, according to which the tail index
measured for each asset and each residue is the same as the tail index of the
Standard \& Poor's 500 index, can be rejected at the 95\% confidence level,
for a given quantile. The values which reject
this hypothesis are indicated by a star in the tables \ref{table:tail2} and
\ref{table:tail3}. During the second time interval from January 1980
to December 2000,
only four residues have a
tail index significantly different from that of that Standard \&
Poor's 500, and only in
the negative tail. The situation is not as good during the first time
interval, especially
for the negative tail, for which not less 13 assets and 10
residues out of 20 have a tail index significantly different from the Standard
\& Poor's 500 ones, for the 5\% quantile.

To summarize, our tests confirm that the tail indexes of most stock return
distributions range between three and four, even though no better precision
can be given with good significance.
Moreover, in most cases, we can assume that both the asset, the factor
and the residue have the same tail index.  We can also add that, as asserted by
\cite{LP94} or \cite{L96}, we cannot reject the hypothesis that the tail index
remains the same over time. Nevertheless, it seems that during the first
period from July 1962 to December 1979,
the tail indexes were sightly larger than during the second period
from January 1980
to december 2000.

\subsection{Determination of the coefficient of tail dependence}

Using the just established empirical fact that
we cannot reject the hypothesis that the
assets, the market and the residues have the same tail index, we can
use the theorem
of Appendix A and its second corollary stated in section
\ref{sec:th}. This leads
us to conclude that one cannot reject the hypothesis of a
non-vanishing tail dependence between the
assets and the market. In addition, the coefficient of tail dependence is
given by equation (\ref{eq:ctd}). In order to determine its value,
we need to estimate the
scale factors for the different assets to derive their coefficient of tail
dependence, according to the formula (\ref{eq:ctd}).

We proceed as follows. Consider a
variable $X$ which asymptotically follows a power law
distribution $\Pr\{X>x\}
\sim C \cdot x^{-\alpha}$. Given a rank ordered sample $x_{1,N} \ge x_{2,N}
\ge \cdots \ge x_{N,N}$, the scale factor $C$ can be consistently
estimated from
the $k$ largest realizations by
\be
\hat C = \frac{k}{N} \cdot (x_{k,N})^{\alpha}~.
\ee
In the tail, this estimator is independent of $k$.
Thus, denoting by $\hat C_{Y}$ and $\hat C_{\epsilon}$ the scale factors of the
factor $Y$ and of the noise $\epsilon$ defined in equation
(\ref{eq:FM}), the estimator of
the coefficient of tail dependence is
\be
\hat \lambda = \frac{1}{1+{\hat \beta}^{-\alpha} \cdot \frac{\hat C_Y}{\hat
C_\epsilon}} = \frac{1}{1+\left( \frac{\epsilon_{k,N}}{{\hat \beta}
\cdot y_{k,N}} \right)^\alpha}~. \label{mgmels}
\ee
Since the tail indices $\alpha$ are impossible to determine with sufficient
accuracy other than saying that the $\alpha$ probably fall in the interval
$3-4$ as we have seen above, our strategy is to determine $\hat
\lambda$ using (\ref{mgmels})
for three different common values $\alpha=3$, $3.5$ and $4$. This
procedure allows
us to test for the sensitivity of the scale factor and therefore of the
tail coefficient with respect to the uncertain value of the tail index.

Table \ref{table:lambda_detail} gives the values of the
coefficients of lower tail dependence over the whole time interval from July
1962 to December 2000, under the assumption that the tail index
$\alpha$ equals $3$. The coefficient of tail dependence
is estimated over the first centile, the first quintile and the first
decile to also test for any possible sensitivity on the tail asymptotics.
For each of these quantiles, the mean values, their standard
deviations and their minimum and maximum values are given.
We first remark that the standard deviation of the tail dependence coefficient
remains small compared with its average value and that the minimum and
maximum values cluster closely around its mean value. This shows that the
coefficient of tail dependence is well-estimated by its mean over a
given quantile. Secondly, we find that these estimated
coefficients of tail dependence exhibit a good stability over the varous
quantiles. These two observations enable us to conclude that the average
coefficient of tail dependence over the first centile is sufficient to provide
a good estimate of the true coefficient of tail dependence.

Tables \ref{table:lambda_62_79}, \ref{table:lambda_80_00} and
\ref{table:lambda_62_00} summarize the different values of the
coefficient of tail dependence for both the positive and the negative
tails, under the assumptions that the tail index $\alpha$ equals $3$,
$3.5$ and $4$ respectively, over the three considered time intervals.
Overall, we find that the coefficients of tail dependence are almost
equal for both the negative and the positive tail and that they are
not very sensitive to the value of the tail index in the interval
considered. More precisely,
during the first time interval from July 1962 to December 1979 (table
\ref{table:lambda_62_79}),
the tail dependence is symetric in both the upper and the lower tail.
During the second time interval from January 1980 to December 2000
and over the whole time interval (tables
\ref{table:lambda_80_00} and \ref{table:lambda_62_00}), the coefficient
of lower tail dependence is slightly but systematically larger than the upper
one. Moreover, since these coefficients of tail
dependence are all less than $1/2$, they decrease when the tail index
$\alpha$ increases and the smaller the coefficient of tail dependence,
the larger the decay.

During the first time interval, most of the coefficients of tail dependence
range between 0.15 and 0.35 in both tails, while during the second time
interval, almost all range between 0.10 and 0.25 in the lower tail and
between 0.10 and 0.20 in the upper one. Thus, the tail dependence is
smaller during the last period than during the first one. This result
is interesting because it associates the smaller
(respectively larger) tail dependence to the second (resp. first)
period of larger (resp. smaller)
volatility, as quantified for instance by the excess kurtosis.

\subsection{Comparison with the historical extremes}

Our determination of the coefficient of tail dependences provides predictions
on the probability that future large moves of stocks may be simultaneous to
large moves of the market. This begs for a check over the available historical
period to determine whether
our estimated coefficients of tail dependence are compatible
with the realized historical extremes.

For this, we consider the ten largest losses of the Standard \&
Poor's 500 index
during the two time sub-intervals\footnote{We do not consider the whole
time interval since the ten largest losses over the whole period
coincide  with the
ten largest ones over the second time subinterval, which would
bias the statistics towards the second time interval.}. Since $\lambda_-$ is by
definition equal to
the probability that a given asset incurs a large loss (say, one of its ten
largest losses) conditional on the occurrence of one
of the ten largest losses of the Standard \& Poor's 500 index,
the probability, for this asset, to undergo $n$ of
its ten largest losses simultaneously with any of the ten largest losses of the
Standard \& Poor's 500 index is given by the binomial law with parameter
$\lambda_-$:
\be
\label{eq:binomial}
P_{\lambda_-}(n)={10 \choose n}~ {\lambda_-}^n ~ (1- \lambda_-)^{(10-n)}.
\ee
We stress that our consideration of only the ten largest drops ensures that the
present test is not embodied in the determination of the tail
dependence coefficient,
which has been determined on a robust procedure over the 1\%, 5\% and
10\% quantiles.
We checked that removing these then largest drops does not modify the
determination
of $\lambda_-$. Our present test can thus be considered as
``out-of-sample,'' in this
sense.

Table \ref{table:extremes} presents, for the two time sub-intervals, the
number of extreme losses among the ten largest losses incured by a given asset
which occured simultaneously with one of the ten largest losses of the standard
\& Poor's 500 index. For each asset, we give the probability of occurence of
such a realisation, according to (\ref{eq:binomial}). We notice that during the
first time interval, only two assets are incompatible, at the 95\% confidence
level, with the value of $\lambda_-$ previously determined: Du Pont (E.I.) de
Nemours \& Co. and Texaco Inc. In constrast, during the second time
interval, four
assets reject the value of $\lambda_-$: Coca Cola Corp., Pepsico Inc.,
Pharmicia Corp. and Texas Instruments Inc.

These results are very encouraging. However, there is a noticeable
systematic bias.
Indeed,  during the first time interval, 17 out of the 20 assets have a
realized number of large losses lower than their expected number (according to
the estimated $\lambda_-$), while during the second time interval, 19 out of
the 20 assets have a realized number of large losses larger than their expected
one. Thus, it seems that during the first time interval the number of large
losses is overestimated by $\lambda_-$ while it is underestimated during the
second time interval.

We propose to
explain the underestimation of the number of large losses between
January 1980 and December 2000 by a possible comonitonicity that
occurred during the
October 1987 crash. Indeed, on October 19, 1987, 12 out of the 20
considered assets incurred their most severe loss, which strongly suggests a
comonotonic effect. Table \ref{table:comonotonic} shows the same results as
in table \ref{table:extremes} but corrected by substracting this comonotonic
effect to the number of large losses. The compatibility between the
number of large
losses and the estimated $\lambda_-$ becomes significantly better
since only  Pepsico Inc.
and Pharmicia Corp. are still rejected, and only 16 assets out of 20 are
underestimated, representing a significant decrease of the bias.

Previous works have shown that,
in period of crashes, the market conditions change, herding
effects may become more important and almost dominant, so that the
market enters an
unusual regime, which can be characterized by outliers present in the
distribution
of drawdowns \cite{JS01}. Our detection of an anomalous comonotonicity
can thus be considered as an independent confirmation of the existence of this
abnormal regime.

Obviously, the overestimation of the number of large losses during
the first time
interval can not be ascribe to the comonotonicity of very large
events, which in fact only occurred once for the Coca-Cola Corp. This
overestimation is probably linked with the low ``volatility'' of the
market during this
period, which can have two effects. The first one is to lead to a less accurate
estimation of the scale factor of the power-law distribution of the assets. The
second one is that a market with smaller volatility produices fewer
large losses.
As a consequence,
the asymptotic regime for which the relation $\Pr\{ X < {F_X}^{-1}(u) | Y <
{F_Y}^{-1}(u) \} \simeq \lambda_-$ holds may not be reached in the sample, and
the number of recorded large losses remain lower than that asymptotically
expected.


\section{Concluding remarks}

We have used the framework offered by factor models in order to derive
a general theoretical expression for the coefficient of tail
dependence between a random
variable and any of its explanatory factor. The coefficient of
tail dependence represents the
probability that an asset incurs a large loss (say), assuming that the market
has also undergone a large loss. We find that factors characterized by
rapidly varying distributions, such as Normal or exponential distributions,
always lead to a vanishing coefficient of tail dependence with other stocks.
In constrast, factors with regularly varying distributions, such as
power-law distributions, can exhibit tail dependence with
other stocks, provided that the
idiosyncratic noise distributions of the corresponding stocks
are not fatter-tailed than the factor.

Applying this general result to individual daily stock returns, we have been
able to estimate the coefficient of tail dependence between the
returns of each stock
and those of the market. This determination of the tail dependence relies only
on the simple estimation of the parameters of the underlying factor model and
on the tail parameters of the distribution of the factor and of the
idiosyncratic
noise of each stock. As a consequence, the two strong advantages of
our approach are the following.
\begin{itemize}
\item[-] The coefficients of
tail dependence are estimated non-parametrically. Indeed, we never specify any
explicit expression of the dependence structure, contrary to most
previous works (see \cite{LS01,MS}  or \cite{P01} for instance);
\item[-] Our theoretical result enables us to estimate an extreme
parameter, not accessible by a direct statistical inference. This is
achieved by the measurement
of parameters whose estimation involves a significant part of the data
with sufficient statistics.
\end{itemize}

Having performed this estimation, we have checked the comptatibility of these
estimated coefficients of tail dependence with the historically
realized extreme
losses observed in the empirical time series. A good agreement is
found, notwithstanding a
slight bias which leads to an overestimate of the occurence of large
events during
the period from July 1962 to December 1979 and to an underestimate
during the time
interval from January 1980 to December 2000.

This bias can be explained by the low volatility of the market during the
first period and by a comonotonicity effect, due to the October 1987 crach,
during the second period. Indeed, from july 1962 to December 1979, the
volatility was so low that the distributions of returns
have probably not sampled their tails sufficiently far the probability
of large conditional losses to be represented by its asymptotic expression
given by the coefficient of tail dependence.
The situation is very different for the
period from january 1980 to December 2000. On October 19, 1987, many assets
incurred their largest loss ever. This is presumably the
manifestation of an `abnormal' regime probably due to herding effects and
irrational behaviors and has been previously characterized as yielding
signatures in the form of outliers in the distribution of drawdowns.

Finally,
the observed lack of stationarity exhibited by the coefficient of
tail dependence
across the two time sub-intervals suggests the importance of going beyond
a stationary view of tail dependence and of studying its dynamics.
This question, which could be of great interest in
the context of the contagion problem, is left for a future work.

Our study has focused on the dependence between different risks. In fact,
our theorem can obviously be applied to extreme temporal
dependences, when the variable follows an autoregressive
process. This should provide an estimate of the
probability that a large loss (respectively gain) is followed by another large
loss (resp. gain) in the following period. Such information is
very interesting in investment and hedging strategies.

\newpage
\appendix

\section{Proof of the theorem}
\label{app:1}
\subsection{Statement}

We consider two random variables $X$ and $Y$, related by the relation
\be
X=\beta \cdot Y+ \epsilon,
\ee
where $\epsilon$ is a random variable independent of $Y$ and $\beta$
a non-random
positive coefficient.

Let $P_Y$ and $F_Y$ denote respectively the density with respect to the
Lebesgue measure and the distribution function of the variable $Y$. Let $F_X$
denotes the distribution function of $X$ and $F_\epsilon$ the distribution
function of $\epsilon$. We state the following theorem:

\begin{theorem}
Assuming that
\begin{itemize}
\item[H0:] The variables $Y$ and $\epsilon$ have distribution functions with
infinite support,

\item[H1:]  For all $x \in [1, \infty)$,
\be
\lim_{t \rightarrow \infty} \frac{t~ P_Y(t x)}{ \bar {F_Y} (t)} = f(x),
\ee

\item[H2:] There are real numbers $t_0 >0$, $\delta>0$ and $A>0$,
such that for
    all $t \ge t_0$ and all $x \ge 1$
\be
\frac{\bar F_Y(t x)}{ \bar F_Y (t)} \le \frac{A}{x^\delta},
\ee

\item[H3:] There is a constant $l \in {\mathbb R_+}$, such that
\be
\lim_{u \rightarrow 1} \frac{{F_X}^{-1}(u)}{{F_Y}^{-1}(u)} = l,
\ee
\end{itemize}
then, the coefficient of (upper) tail dependence of $(X,Y)$ is given by
\be
\lambda = \int_{\max\{1, \frac{l}{\beta}\}} ^\infty dx~ f(x).
\ee
\end{theorem}

\subsection{Proof}

We first give a general expression for the probability for $X$ to be
larger than
$F_X^{-1}(u)$ knowing that $Y$ is larger than $ F_Y^{-1}(u)$~:
\begin{lemma}
\label{lem:1}
The probability that $X$ is larger than $F_X^{-1}(u)$ knowing that $Y$ is
larger than $ F_Y^{-1}(u)$ is given by~:
\be
\Pr \left[ X > F_X^{-1}(u) | Y > F_Y^{-1}(u) \right] =
   \frac{F_Y^{-1}(u)}{1-u}  \int_1^{\infty}
dx ~ P_Y \left( F_Y^{-1}(u)~x \right) \cdot
\bar F_\epsilon \left[{F_X}^{-1}(u)-\beta F_Y^{-1}(u)~x \right] ~.
\ee
\end{lemma}

Proof :
\bea
\Pr\{ X> {F_X}^{-1}(u), Y> {F_Y}^{-1}(u) \}  &=& \E \left[ 1_{\{X>
{F_X}^{-1}(u)\}} \cdot 1_{\{Y>{F_Y}^{-1}(u)\}} \right] \\
&=&  \E \left[ \E \left[ 1_{\{X>
{F_X}^{-1}(u)\}} \cdot 1_{\{Y>{F_Y}^{-1}(u)\}} | Y \right] \right]\\
&=&  \E \left[1_{\{Y>{F_Y}^{-1}(u)\}} \cdot \E \left[ 1_{\{X>
{F_X}^{-1}(u)\}}  | Y \right] \right]\\
&=&  \E \left[1_{\{Y>{F_Y}^{-1}(u)\}} \cdot \E \left[ 1_{\{\epsilon >
{F_X}^{-1}(u) - \beta Y\}} \right] \right]\\
&=&  \E \left[1_{\{Y>{F_Y}^{-1}(u)\}} \cdot \bar F_\epsilon (
{F_X}^{-1}(u) - \beta Y) \right]
\eea

Assuming that the variable $Y$ admits a density $P_Y$ with respect to the
Lebesgue measure, this yields
\be
\Pr\{ X> {F_X}^{-1}(u), Y> {F_Y}^{-1}(u) \}  = \int_{F_Y^{-1}(u)}^{\infty}
dy ~ P_Y(y) \cdot
\bar F_\epsilon [{F_X}^{-1}(u)-\beta y]~.
\ee
Performing the change of variable $ x= {F_Y}^{-1}(u) \cdot y$, in
the equation above, we obtain
\be
\Pr\{ X> {F_X}^{-1}(u), Y> {F_Y}^{-1}(u) \}  =F_Y^{-1}(u)~ \int_1^{\infty}
dx ~ P_Y(F_Y^{-1}(u)~x) \cdot
\bar F_\epsilon [{F_X}^{-1}(u)-\beta F_Y^{-1}(u)~x]~,
\ee

and, dividing by $\bar {F_Y} \left( {F_Y}^{-1}(u) \right) = 1- u$, this
concludes the proof.
$\square$

Let us now define the function
\be
f_u(x) =  \frac{F_Y^{-1}(u)}{1-u} ~ P_Y(F_Y^{-1}(u)~x) \cdot
\bar F_\epsilon [{F_X}^{-1}(u)-\beta F_Y^{-1}(u)~x]~.
\ee
We can state the following result
\begin{lemma}
Under assumption $H1$ and $H3$, for all $x \in [1, \infty)$,
\be
f_u(x) \longrightarrow 1_{\left\{x> \frac{l}{\beta}\right\}} \cdot f(x),
\ee
almost everywhere, as u goes to 1.
\end{lemma}

Proof: Let us apply the assumption $H1$. We have
\bea
\lim_{u \rightarrow 1} \frac{F_Y^{-1}(u)}{1-u} ~ P_Y(F_Y^{-1}(u)~x) &=&
\lim_{t \rightarrow \infty}   \frac{t~ P_Y(t~x)}{\bar{F_Y}(t)} ~,\\
&=& f(x).
\eea
Applying now the assumption $H3$, we have
\bea
\lim_{u\rightarrow 1}{F_X}^{-1}(u)-\beta F_Y^{-1}(u)~x
&=&  \lim_{u\rightarrow 1}  \beta F_Y^{-1}(u)~\left(
\frac{{F_X}^{-1}(u)}{\beta F_Y^{-1}(u)~}-x \right)\\
&=& \left\{
\begin{array}{cc}
-\infty&\mbox{if}~x>\frac{l}{\beta},\\
\infty&\mbox{if}~x<\frac{l}{\beta},
\end{array}
\right.  \\
\eea
which gives
\be
\lim_{u\rightarrow 1} \bar F_\epsilon [{F_X}^{-1}(u)-\beta F_Y^{-1}(u)~x]
= 1_{\left\{x > \frac{l}{\beta}\right\}},
\ee
and finally
\bea
\lim_{u \rightarrow 1} f_u(x) &=& \lim_{u \rightarrow 1}
\frac{F_Y^{-1}(u)}{1-u} ~ P_Y(F_Y^{-1}(u)~x) \cdot \lim_{u\rightarrow 1} \bar
F_\epsilon [{F_X}^{-1}(u)-\beta F_Y^{-1}(u)~x] ,\\
&=& 1_{\left\{x > \frac{l}{\beta}\right\}} \cdot f(x),
\eea
which concludes the proof.
$\square$

Let us now proove that there exists an integrable function $g(x)$ such
that, for all $t \ge t_0$ and all $x \ge 1$, we have $f_t (x) \le g(x)$.
Indeed, let us write
\be
\frac{ t ~ P_Y (t x)}{ \bar F_Y(t)} = \frac{ t ~ P_Y (t x)}{
   \bar F_Y(t x)} \cdot \frac{\bar F_Y(t x)}{\bar F_Y(t)}~.
   \label{mhmhtle}
\ee
For the leftmost factor in the right-hand-side of equation
(\ref{mhmhtle}), we easily
obtain
\be
\forall t, ~ \forall x \ge 1, ~~~~ \frac{ t ~ P_Y (t x)}{
   \bar F_Y(t x)} \le \frac{x^* ~P_Y(x^*)}{\bar F_Y (x^*)} \cdot
\frac{1}{x},
\ee
where $x^*$ denotes the point where the function
$\frac{x~P_Y(x)}{\bar F_Y(x)}$ reaches its maximum. The rightmost
factor in the right-hand-side of (\ref{mhmhtle})
is smaller than $A/x^\delta$ by assumption $H2$, so that
\be
\forall t \ge t_0, ~\forall x \ge 1, ~~~~ \frac{ t ~ P_Y (t x)}{
   \bar F_Y(t)} \le \frac{x^* ~P_Y(x^*)}{\bar F_Y (x^*)} \cdot
\frac{A}{x^{1+\delta}}~.
\ee

Posing
\be
g(x)=\frac{x^* ~P_Y(x^*)}{\bar F_Y (x^*)} \cdot
\frac{A}{x^{1+\delta}}~,
\ee
and recalling that, for all $\epsilon \in {\mathbb R}$, $\bar F_\epsilon
(\epsilon) \le 1$,
we have found an integrable function such that for some $u_0 \ge 0$, we have
\be
\forall u \in [u_0, 1), ~\forall x \ge 1, ~~~~ f_u(x) \le g(x)~.
\ee

Thus, applying Lebesgue's theorem of dominated convergence, we can assert that
\be
\lim_{u \rightarrow 1} \int_1^\infty dx~ f_u(x) = \int_1^\infty dx ~
1_{\left\{x > \frac{l}{\beta}\right\}} \cdot f(x).
\ee

Since
\bea
\lim_{u \rightarrow 1} \int_1^\infty dx~ f_u(x) &=&  \lim_{u \rightarrow 1} \Pr
\left[ X > F_X^{-1}(u) | Y > F_Y^{-1}(u) \right], \\
&=& \lambda,
\eea
the proof of theorem 1 is concluded.
$\square$

\section{Proofs of the corollaries}
\subsection{First corollary}
\label{app:c1}

\begin{corollary}

If the random variable $Y$ has a rapidly varying distribution function, then
$\lambda = 0$.

\end{corollary}

Proof : Let us write
\be
\label{eq:c1}
\frac{t~ P_Y(t x)}{ \bar {F_Y} (t)}  = \frac{t~ P_Y(t x)}{ \bar {F_Y}
(t x)} \cdot \frac{\bar {F_Y} (t x)}{\bar {F_Y} (t)}.
\ee
For a rapidly varying function $\bar F_Y$, we have
\be
\forall x >1, ~~~\lim_{t \rightarrow \infty} \frac{\bar F_Y(t
x)}{\bar F_Y(t)} = 0,
\ee
while the leftmost factor of the right-hand-side of equation
(\ref{eq:c1}) remains bounded
as $t$ goes to infinity, so that
\be
\lim_{t \rightarrow \infty} \frac{t~ P_Y(t x)}{ \bar {F_Y}
(t x)} \cdot \frac{\bar {F_Y} (t x)}{\bar {F_Y} (t)} = f(x) =0~.
\ee
Since $f(x)=0$, we can apply lemma 2 without  the hypothesis $H3$, which
concludes the proof.
$\square$

\subsection{Second corollary}
\label{app:c2}

\begin{corollary}
Let $Y$ be regularly varying with index $(-\alpha)$, and assume that hypothesis
$H3$ is satisfied. Then, the coefficient of (upper) tail dependence is
\be
\lambda =\frac{1}{ \left[ \max \left\{1, \frac{l}{\beta} \right\}
\right]^\alpha},
\ee
where $l$ denotes the limit, when $u \to 1$, of the ratio
${F_X}^{-1}(u)/{F_Y}^{-1}(u)$.
\end{corollary}

Proof : Karamata's theorem (see \cite[p 567]{Embrechtsbook}) ensures that
$H1$ is satisfied with $f(x) = \frac{\alpha}{x^{\alpha+1}}$, which is 
sufficient
to proove the corollary. To go one step further, let us define
\bea
\bar F_y(y) &=& y^{-\alpha} \cdot L_1(y), \\
\bar F_\epsilon(\epsilon) &=& \epsilon^{-\alpha} \cdot L_2(\epsilon),
\eea
where $L_1(\cdot)$ and $L_2(\cdot)$ are slowly varying functions.

Using the proposition stated in \cite[p 278]{Feller_II}, we obtain, for the
distribution of the variable $X$
\be
\bar F_X(x) \sim x^{-\alpha} \left( \beta^\alpha \cdot L_1 \left(
\frac{x}{\beta} \right) + L_2(x) \right),
\ee
for large $x$.

Assuming now, for simplicity, that $L_1$ (resp. $L_2$) goes to a
constant $C_1$ (resp. $C_2$), this implies that $H3$ is satistified, since
\be
l=\lim_{u \rightarrow 1} \frac{{F_X}^{-1}(u)}{{F_Y}^{-1}(u)} = \beta \left[ 1+
\frac{C_2}{\beta^\alpha~ C_1} \right]^\frac{1}{\alpha}~.
\ee
This allows us to obtain the equations (\ref{eq:8}) and (\ref{eq:ctd}).
$\square$


\newpage

\begin{landscape}


\begin{table}
\begin{center}

{\bf Statistical description of the set of studied stocks}

\vspace{1cm}

\begin{tabular}{lcccrccccrccccr}
\hline
 & \multicolumn{4}{c}{July 1962 - December 1979 } &  & \multicolumn{4}{c}{January 1980 - December 2000 } &  & \multicolumn{4}{c}{July 1962 - December 2000 } \\
\cline{2-5}
\cline{7-10}
\cline{12-15}
 & Mean & Std. & Skew. & Kurt. &  & Mean & Std. & Skew. & Kurt. &  & Mean & Std. & Skew. & Kurt. \\
\hline
Abbott Labs & 0.6677 & 0.0154 & 0.2235 & 2.192 &  & 0.9217 & 0.0174 & -0.0434 & 2.248 &  & 0.8066 & 0.0165 & 0.0570 & 2.300\\
American Home Products Corp. & 0.4755 & 0.0136 & 0.2985 & 3.632 &  & 0.8486 & 0.0166 & 0.1007 & 8.519 &  & 0.6803 & 0.0154 & 0.1717 & 7.557 \\
Boeing Co. & 0.8460 & 0.0228 & 0.6753 & 4.629 &  & 0.7752 & 0.0193 & 0.1311 & 4.785 &  & 0.8068 & 0.0209 & 0.4495 & 4.901 \\
Bristol-Myers Squibb Co. & 0.5342 & 0.0152 & -0.0811 & 2.808 &  & 0.9353 & 0.0175 & -0.3437 & 16.733 &  & 0.7546 & 0.0165 & -0.2485 & 12.573 \\
Chevron Corp. & 0.4916 & 0.0134 & 0.2144 & 2.442 &  & 0.6693 & 0.0169 & 0.0491 & 4.355 &  & 0.5885 & 0.0154 & 0.1033 & 4.209 \\
Du Pont (E.I.) de Nemours \& Co. & 0.2193 & 0.0126 & 0.3493 & 2.754 &  & 0.6792 & 0.0172 & -0.1021 & 4.731 &  & 0.4715 & 0.0153 & 0.0231 & 4.937 \\
Disney (Walt) Co. & 0.9272 & 0.0215 & 0.2420 & 2.762 &  & 0.8759 & 0.0195 & -0.6661 & 17.655 &  & 0.8997 & 0.0204 & -0.1881 & 9.568 \\
General Motors Corp. & 0.3547 & 0.0126 & 0.4138 & 4.302 &  & 0.5338 & 0.0183 & -0.0128 & 5.373 &  & 0.4538 & 0.0160 & 0.0872 & 6.164 \\
Hewlett-Packard Co. & 0.7823 & 0.0199 & 0.0212 & 3.063 &  & 0.8913 & 0.0238 & 0.0254 & 4.921 &  & 0.8420 & 0.0221 & 0.0256 & 4.624 \\
Coca-Cola Co. & 0.4829 & 0.0138 & 0.0342 & 5.436 &  & 0.9674 & 0.0170 & -0.1012 & 14.377 &  & 0.7483 & 0.0157 & -0.0513 & 12.611 \\
Minnesota Mining \& MFG Co. & 0.3459 & 0.0139 & 0.3016 & 2.997 &  & 0.6885 & 0.0150 & -0.7861 & 20.609 &  & 0.5333 & 0.0145 & -0.3550 & 14.066 \\
Philip Morris Cos Inc. & 0.7930 & 0.0153 & 0.2751 & 2.799 &  & 0.9664 & 0.0180 & -0.2602 & 10.954 &  & 0.8863 & 0.0169 & -0.0784 & 8.790 \\
Pepsico Inc. & 0.4982 & 0.0147 & 0.2380 & 2.867 &  & 0.9443 & 0.0180 & 0.1372 & 4.594 &  & 0.7431 & 0.0166 & 0.1786 & 4.413 \\
Procter \& Gamble Co. & 0.3569 & 0.0115 & 0.3911 & 4.343 &  & 0.7916 & 0.0164 & -1.6610 & 46.916 &  & 0.5947 & 0.0144 & -1.2408 & 44.363 \\
Pharmacia Corp. & 0.3801 & 0.0145 & 0.2699 & 3.508 &  & 0.9027 & 0.0191 & -0.6133 & 13.587 &  & 0.6666 & 0.0172 & -0.3773 & 12.378 \\
Schering-Plough Corp. & 0.6328 & 0.0163 & 0.2619 & 3.112 &  & 1.0663 & 0.0192 & 0.1781 & 7.9979&  & 0.8703 & 0.0179 & 0.2139 & 6.757 \\
Texaco Inc. & 0.3416 & 0.0134 & 0.2656 & 2.596 &  & 0.6644 & 0.0166 & 0.1192 & 6.477 &  & 0.5197 & 0.0152 & 0.1725 & 5.829 \\
Texas Instruments Inc. & 0.6839 & 0.0198 & 0.2076 & 3.174 &  & 1.0299 & 0.0268 & 0.1595 & 7.848 &  & 0.8726 & 0.0239 & 0.1831 & 7.737 \\
United Technologies Corp & 0.5801 & 0.0185 & 0.3397 & 2.826 &  & 0.7752 & 0.0170 & 0.0396 & 3.190 &  & 0.6876 & 0.0177 & 0.1933 & 3.034 \\
Walgreen Co. & 0.5851 & 0.0165 & 0.3530 & 3.030 &  & 1.1996 & 0.0185 & 0.1412 & 3.316 &  & 0.9217 & 0.0176 & 0.2260 & 3.295 \\
 &  &  &  &  &  &  &  &  &  &  &  &  &  &  \\
Standart \& Poor's 500 & 0.1783 & 0.0075 & 0.2554 & 3.131 &  & 0.5237 & 0.0101 & -1.6974 & 36.657 &  & 0.3674 & 0.0090 & -1.2236 & 32.406\\
\hline
\end{tabular}

\caption{\label{table:asset} This table gives the main statistical features of
the three samples we have considered. The columns {\it Mean, Std., Skew.} and
{\it Kurt.} respectively give the average return multiplied by one thousand,
the standard deviation, the skewness and the excess kurtosis of each
asset over the
time intervals form July 1962 to December 1979, January 1980 to Decemeber 2000
and July 1962 to December 2000. The excess kurtosis is given as
indicative of the
relative weight of large return amplitudes, and can always be calculated
over a finite time series even if
it may not be asymptotically defined for power tails with exponents
less than $4$.
}

\end{center}
\end{table}


\begin{table}
\begin{center}

{\bf Estimation of the parameters of the factor model}

\vspace{1cm}

\begin{tabular}{lcccccccc}
\hline
 & \multicolumn{2}{c}{July 1962 - December 1979 } &  & \multicolumn{2}{c}{January 1979 - December 2000 } &  & \multicolumn{2}{c}{July 1962 - December 1979 } \\
\cline{2-3}
\cline{5-6}
\cline{8-9}
 & $\beta$ & $\rho$ &  & $\beta$ & $\rho$ &  & $\beta$ & $\rho$ \\
\hline
Abbott Labs & 0.9010 & -0.0009 &  & 0.9145 & -0.0016 &  & 0.9103 & -0.0013 \\
American Home Products Corp. & 0.9865 & -0.0006 &  & 0.8124 & -0.0015 &  & 0.8668 & -0.0011 \\
Boeing Co. & 1.4435 & -0.0007 &  & 0.9052 & -0.0009 &  & 1.0733 & -0.0009 \\
Bristol-Myers Squibb Co. & 1.0842 & -0.0006 &  & 1.0455 & -0.0014 &  & 1.0576 & -0.0011 \\
Chevron Corp. & 1.0072 & -0.0007 &  & 0.8345 & -0.0008 &  & 0.8885 & -0.0008 \\
Du Pont (E.I.) de Nemours \& Co. & 1.0819 & -0.0001 &  & 0.9461 & -0.0007 &  & 0.9885 & -0.0004 \\
Disney (Walt) Co. & 1.5551 & -0.0009 &  & 1.0034 & -0.0011 &  & 1.1757 & -0.0011 \\
General Motors Corp. & 1.0950 & -0.0004 &  & 1.0112 & 0.0000 &  & 1.0374 & -0.0002 \\
Hewlett-Packard Co. & 1.3926 & -0.0008 &  & 1.3085 & -0.0005 &  & 1.3348 & -0.0008 \\
Coca-Cola Co. & 1.0357 & -0.0006 &  & 0.9856 & -0.0017 &  & 1.0012 & -0.0012 \\
Minnesota Mining \& MFG Co. & 1.1344 & -0.0003 &  & 0.8768 & -0.0010 &  & 0.9573 & -0.0006 \\
Philip Morris Cos Inc. & 1.0913 & -0.0011 &  & 0.8624 & -0.0017 &  & 0.9339 & -0.0015 \\
Pepsico Inc. & 0.9597 & -0.0006 &  & 0.9028 & -0.0016 &  & 0.9206 & -0.0011 \\
Procter \& Gamble Co. & 0.8299 & -0.0005 &  & 0.8955 & -0.0012 &  & 0.8750 & -0.0009 \\
Pharmacia Corp. & 1.0756 & -0.0004 &  & 0.8846 & -0.0013 &  & 0.9443 & -0.0009 \\
Schering-Plough Corp. & 1.1258 & -0.0007 &  & 1.0506 & -0.0017 &  & 1.0741 & -0.0013 \\
Texaco Inc. & 1.4592 & -0.0006 &  & 1.3826 & -0.0007 &  & 1.4065 & -0.0007 \\
Texas Instruments Inc. & 0.9419 & -0.0004 &  & 0.6617 & -0.0011 &  & 0.7492 & -0.0007 \\
United Technologies Corp & 1.1348 & -0.0005 &  & 0.9064 & -0.0011 &  & 0.9777 & -0.0009 \\
Walgreen Co. & 0.6369 & -0.0007 &  & 0.8592 & -0.0024 &  & 0.7898 & -0.0016 \\
\hline
\end{tabular}

\caption{\label{table:param} This table presents the estimated coefficient
$\beta$ for the factor model (\ref{eq:FM}) and the correlation coefficient
$\rho$ between the factor and the estimated idiosyncratic noise, for the
different time intervals we have considered.  A Fisher's test shows that at the
95\% confidence level none of the correlation coefficient is significantly
different from zero.}

\end{center}
\end{table}


\begin{table}
\begin{center}

{\bf Tail index for the time interval from July 1962 to December 1979}
\vspace{1cm}

\begin{tabular}{lccccccccccccccccc}
\hline
 & \multicolumn{8}{c}{Negative Tail} &  & \multicolumn{8}{c}{Positive Tail} \\
\cline{2-9}
\cline{11-18}
 & \multicolumn{2}{c}{q = 1\%} &  & \multicolumn{2}{c}{q = 2.5\%} &  & \multicolumn{2}{c}{q = 5\%} &  & \multicolumn{2}{c}{q = 1\%} &  & \multicolumn{2}{c}{q = 2.5\%} &  & \multicolumn{2}{c}{q = 5\%} \\
\cline{2-3}
\cline{5-6}
\cline{8-9}
\cline{11-12}
\cline{14-15}
\cline{17-18}
 & Asset & $\epsilon$ &  & Asset & $\epsilon$ &  & Asset & $\epsilon$ &  & Asset & $\epsilon$ &  & Asset & $\epsilon$ &  & Asset & $\epsilon$ \\
\hline
Abbott Labs & 5.54 & 5.31 &  & 3.94 & 4.02 &  & 3.27 & 3.31 &  & 5.10 & 4.50 &  & 4.09 & 3.71 &  & 3.53$^*$ & 3.14 \\
American Home Products Corp. & 4.58 & 5.11 &  & 3.89 & 3.81 &  & 3.02$^*$ & 3.21$^*$ &  & 3.64 & 4.66 &  & 3.60 & 3.81 &  & 3.11 & 3.15 \\
Boeing Co. & 6.07 & 4.90 &  & 4.57 & 3.74 &  & 3.32 & 3.49 &  & 4.04 & 4.27 &  & 3.95 & 4.19 &  & 3.35$^*$ & 2.93 \\
Bristol-Myers Squibb Co. & 4.32 & 4.27 &  & 3.31 & 3.95 &  & 2.99$^*$ & 3.16$^*$ &  & 5.96$^*$ & 5.19 &  & 3.94 & 4.82$^*$ &  & 3.62$^*$ & 4.03$^*$ \\
Chevron Corp. & 5.24 & 4.78 &  & 3.75 & 3.29 &  & 2.91 & 3.12$^*$ &  & 5.21 & 5.15 &  & 3.90 & 4.26 &  & 3.25$^*$ & 3.07 \\
Du Pont (E.I.) de Nemours \& Co. & 5.26 & 4.36 &  & 3.69 & 3.76 &  & 3.17$^*$ & 3.23$^*$ &  & 5.35 & 5.15 &  & 4.00 & 3.37 &  & 3.13 & 3.04 \\
Disney (Walt) Co. & 3.59 & 4.23 &  & 3.59 & 3.84 &  & 3.08$^*$ & 3.22$^*$ &  & 4.90 & 4.34 &  & 4.26 & 3.73 &  & 3.33$^*$ & 3.29$^*$ \\
General Motors Corp. & 4.82 & 3.50 &  & 3.72 & 3.66 &  & 2.94$^*$ & 3.36 &  & 3.91 & 4.78 &  & 3.64 & 3.86 &  & 2.94 & 3.07 \\
Hewlett-Packard Co. & 3.76 & 3.89 &  & 3.12$^*$ & 3.05$^*$ &  & 2.81$^*$ & 3.00$^*$ &  & 4.64 & 5.08 &  & 4.08 & 4.20 &  & 3.41$^*$ & 3.42$^*$ \\
Coca-Cola Co. & 3.45 & 3.45 &  & 3.05$^*$ & 3.71 &  & 2.75$^*$ & 3.17$^*$ &  & 3.91 & 4.26 &  & 3.16 & 3.61 &  & 2.81 & 3.16 \\
Minnesota Mining \& MFG Co. & 5.16 & 4.86 &  & 4.06 & 4.35 &  & 3.43 & 3.71 &  & 4.35 & 4.47 &  & 3.96 & 3.31 &  & 3.14 & 3.06 \\
Philip Morris Cos Inc. & 4.63 & 3.79 &  & 3.82 & 3.90 &  & 3.38 & 3.48 &  & 4.10 & 4.64 &  & 3.59 & 3.85 &  & 3.03 & 3.06 \\
Pepsico Inc. & 4.89 & 5.35 &  & 3.93 & 4.49 &  & 3.02$^*$ & 3.27 &  & 4.07 & 4.67 &  & 3.49 & 3.86 &  & 3.15 & 3.21$^*$ \\
Procter \& Gamble Co. & 4.42 & 3.77 &  & 3.77 & 3.74 &  & 3.13$^*$ & 3.42 &  & 4.14 & 5.39 &  & 3.59 & 3.73 &  & 2.97 & 3.40$^*$ \\
Pharmacia Corp. & 4.73 & 4.24 &  & 4.05 & 3.45 &  & 2.88$^*$ & 3.34 &  & 4.46 & 3.72 &  & 3.95 & 3.90 &  & 3.14 & 2.99 \\
Schering-Plough Corp. & 4.59 & 4.70 &  & 4.20 & 3.87 &  & 3.37 & 3.33 &  & 4.60 & 5.88$^*$ &  & 3.50 & 3.91 &  & 3.07 & 3.22$^*$ \\
Texaco Inc. & 5.34 & 4.59 &  & 3.99 & 3.84 &  & 3.07$^*$ & 3.19$^*$ &  & 3.83 & 4.10 &  & 3.94 & 3.67 &  & 3.14 & 2.98 \\
Texas Instruments Inc. & 4.08 & 4.54 &  & 3.36 & 3.13$^*$ &  & 3.22$^*$ & 2.87$^*$ &  & 4.52 & 4.20 &  & 3.67 & 3.79 &  & 3.16 & 3.07 \\
United Technologies Corp & 4.00 & 4.49 &  & 3.52 & 3.92 &  & 3.27 & 3.46 &  & 4.78 & 4.97 &  & 3.73 & 3.98 &  & 3.26$^*$ & 3.49$^*$ \\
Walgreen Co. & 4.63 & 6.50 &  & 3.85 & 4.26 &  & 2.94$^*$ & 3.18$^*$ &  & 5.16 & 4.56 &  & 3.47 & 3.30 &  & 3.15 & 2.82 \\
 &  &  &  &  &  &  &  &  &  &  &  &  &  &  &  &  &  \\
Standart \& Poor's 500 & 5.17 & - &  & 4.16 & - &  & 3.91 & - &  & 3.74 & - &
& 3.34 & - &  & 2.64 & - \\ \hline
\end{tabular}

\caption{\label{table:tail2} This table gives the estimated value of the tail
index for the twenty considered assets, the Standard \& Poor's 500 index and
the residues obtained by regressing each asset on the Standard \& Poor's 500
index, for both the negative and the positive tails, during the time interval
from July 1962 to December 1979. The tail indexes are estimed by the Hill's
estimator at the quantile 1\%, 2.5\% and 5\% which are the optimal quantiles
given by the \cite{H90} and \cite{DdV97}'s algorithms. The values
decorrated with stars
represent the tail indexes which cannot be considered equal to the
Standard \& Poor's 500 index's tail index at the 95\% confidence level.}

\end{center}
\end{table}

\begin{table}
\begin{center}

{\bf Tail index for the time interval from January 1980 to December 2000}
\vspace{1cm}

\begin{tabular}{lccccccccccccccccc}
\hline
 & \multicolumn{8}{c}{Negative Tail} &  & \multicolumn{8}{c}{Positive Tail} \\
\cline{2-9}
\cline{11-18}
 & \multicolumn{2}{c}{q = 1\%} &  & \multicolumn{2}{c}{q = 2.5\%} &  & \multicolumn{2}{c}{q = 5\%} &  & \multicolumn{2}{c}{q = 1\%} &  & \multicolumn{2}{c}{q = 2.5\%} &  & \multicolumn{2}{c}{q = 5\%} \\
\cline{2-3}
\cline{5-6}
\cline{8-9}
\cline{11-12}
\cline{14-15}
\cline{17-18}
 & Asset & $\epsilon$ &  & Asset & $\epsilon$ &  & Asset & $\epsilon$ &  & Asset & $\epsilon$ &  & Asset & $\epsilon$ &  & Asset & $\epsilon$ \\
\hline
Abbott Labs & 3.59 & 3.60 &  & 3.35 & 3.62 &  & 3.22 & 3.39 &  & 5.14 & 4.60 &  & 4.16 & 3.76 &  & 3.77 & 3.07 \\
American Home Products Corp. & 3.03 & 3.07 &  & 3.11 & 2.78 &  & 2.73 & 2.49$^*$ &  & 4.01 & 3.47 &  & 3.28 & 3.02 &  & 2.87 & 2.79 \\
Boeing Co. & 3.39 & 3.97 &  & 3.23 & 3.53 &  & 3.02 & 3.21 &  & 4.86 & 3.65 &  & 3.45 & 3.16 &  & 3.13 & 3.23 \\
Bristol-Myers Squibb Co. & 3.21 & 3.15 &  & 2.90 & 3.41 &  & 2.80 & 3.16 &  & 2.98 & 3.74 &  & 3.35 & 3.12 &  & 3.20 & 2.75 \\
Chevron Corp. & 4.13 & 4.48 &  & 3.99 & 3.91 &  & 3.30 & 3.45 &  & 5.16 & 4.53 &  & 3.88 & 3.81 &  & 3.01 & 3.06 \\
Du Pont (E.I.) de Nemours \& Co. & 3.99 & 3.49 &  & 3.76 & 3.23 &  & 3.02 & 3.04 &  & 5.36 & 4.33 &  & 4.31 & 3.35 &  & 3.44 & 2.76 \\
Disney (Walt) Co. & 2.83 & 3.24 &  & 2.76 & 2.97 &  & 2.85 & 2.83 &  & 3.97 & 3.70 &  & 3.68 & 3.33 &  & 3.15 & 2.87 \\
General Motors Corp. & 4.44 & 4.79 &  & 3.88 & 4.27$^*$ &  & 3.44 & 3.56 &  & 5.76 & 5.32 &  & 4.45 & 3.86 &  & 3.43 & 3.22 \\
Hewlett-Packard Co. & 3.73 & 3.45 &  & 3.52 & 3.12 &  & 3.00 & 2.73 &  & 4.31 & 3.40 &  & 3.47 & 3.29 &  & 3.24 & 2.99 \\
Coca-Cola Co. & 3.01 & 3.76 &  & 3.14 & 3.48 &  & 2.99 & 2.86 &  & 4.06 & 3.47 &  & 3.45 & 3.16 &  & 3.37 & 2.87 \\
Minnesota Mining \& MFG Co. & 3.52 & 3.38 &  & 3.21 & 3.39 &  & 2.88 & 3.04 &  & 3.76 & 3.46 &  & 3.95 & 3.22 &  & 3.10 & 2.76 \\
Philip Morris Cos Inc. & 3.58 & 3.34 &  & 3.33 & 3.12 &  & 2.68 & 2.53$^*$ &  & 3.42 & 3.16 &  & 3.70 & 3.07 &  & 2.85 & 2.81 \\
Pepsico Inc. & 4.14 & 4.46 &  & 3.39 & 3.60 &  & 2.99 & 3.27 &  & 4.00 & 3.87 &  & 3.61 & 3.34 &  & 3.44 & 3.31 \\
Procter \& Gamble Co. & 2.65 & 2.46 &  & 3.29 & 3.19 &  & 3.19 & 2.87 &  & 4.35 & 3.90 &  & 3.48 & 3.20 &  & 3.14 & 2.91 \\
Pharmacia Corp. & 2.96 & 3.20 &  & 3.09 & 2.79 &  & 2.80 & 2.70 &  & 4.12 & 4.70 &  & 3.44 & 3.50 &  & 3.31 & 2.89 \\
Schering-Plough Corp. & 4.22 & 5.20$^*$ &  & 3.29 & 3.68 &  & 3.11 & 3.05 &  & 3.23 & 3.51 &  & 3.45 & 3.08 &  & 3.06 & 2.87 \\
Texaco Inc. & 3.09 & 3.20 &  & 3.10 & 3.15 &  & 2.88 & 2.84 &  & 3.65 & 3.36 &  & 3.20 & 3.04 &  & 2.86 & 2.70 \\
Texas Instruments Inc. & 3.49 & 3.53 &  & 3.35 & 3.31 &  & 2.89 & 2.99 &  & 4.00 & 3.42 &  & 3.36 & 3.30 &  & 2.97 & 3.06 \\
United Technologies Corp & 4.21 & 3.98 &  & 3.82 & 3.46 &  & 3.34 & 3.18 &  & 5.39 & 4.50 &  & 4.00 & 3.80 &  & 3.51 & 3.26 \\
Walgreen Co. & 4.06 & 4.35 &  & 3.81 & 4.04 &  & 3.20 & 3.40 &  & 4.60 & 5.12 &  & 3.79 & 3.54 &  & 3.20 & 3.07 \\
 &  &  &  &  &  &  &  &  &  &  &  &  &  &  &  &  &  \\
Standart \& Poor's 500 & 3.16 & - &  & 3.17 & - &  & 3.16 & - &  & 4.00 & - &
& 3.65 & - &  & 3.19 & - \\ \hline
\end{tabular}

\caption{\label{table:tail3} This table gives the estimated value of the tail
index for the twenty considered assets, the Standard \& Poor's 500 index and
the residues obtained by regressing each asset on the Standard \& Poor's 500
index, for both the negative and the positive tails, during the time interval
from January 1980 to December 2000. The tail indexes are estimed by the Hill's
estimator at the quantile 1\%, 2.5\% and 5\% which are the optimal quantiles
given by the \cite{H90} and \cite{DdV97}'s algorithms. The values
decorrated with stars
represent the tail indexes whose value cannot be considered equal to the
Standard \& Poor's 500 index's tail index at the 95\% confidence level.}

\end{center}
\end{table}


\begin{table}
\begin{center}

{\bf Coefficient of lower tail dependence during the time interval from July
1962 to December 2000 for a tail index equal to three}
\vspace{1cm}

\begin{tabular}{lcccccccccccccc}

\hline
&\multicolumn{4}{c}{First Centile}&&\multicolumn{4}{c}{First Quintile}&&\multicolumn{4}{c}{First Decile}\\
\cline{2-5}
\cline{7-10}
\cline{12-15}
&mean&std.&min.&max.&&mean&std.&min.&max.&&mean&std.&min.&max.\\
\hline

Abbott Labs&0.1670&0.0127&0.1442&0.2137&&0.1633&0.0071&0.1442&0.2137&&0.1540&0.0120&0.1331&0.2137\\
American Home Products Corp.&0.1423&0.0207&0.0910&0.1720&&0.1728&0.0205&0.091&0.1963&&0.1823&0.0175&0.0910&0.2020 \\
Boeing Co.&0.1372&0.0127&0.1101&0.1804&&0.1349&0.0064&0.1101&0.1804&&0.1289&0.0078&0.1101&0.1804 \\
Bristol-Myers Squibb Co.&0.2720&0.0231&0.1878&0.3052&&0.2751&0.0115&0.1878&0.3052&&0.2696&0.0110&0.1878&0.3052\\
Chevron Corp.&0.1853&0.0188&0.1656&0.2564&&0.1790&0.0105&0.1634&0.2564&&0.1748&0.0096&0.1606&0.2564\\
Du Pont (E.I.) de Nemours \& Co.&0.2547&0.0148&0.2127&0.2871&&0.2695&0.0117&0.2127&0.2876&&0.2685&0.0103&0.2127&0.2876 \\
Disney (Walt) Co.&0.1772&0.0149&0.1368&0.1957&&0.1938&0.0123&0.1368&0.2094&&0.1900&0.0109&0.1368&0.2094 \\
General Motors Corp.&0.2641&0.0259&0.2393&0.3652&&0.2565&0.0138&0.2349&0.3652&&0.2545&0.0108&0.2349&0.3652 \\
Hewlett-Packard Co.&0.1701&0.0096&0.1389&0.1914&&0.2018&0.0230&0.1389&0.2303&&0.2039&0.0176&0.1389&0.2303 \\
Coca-Cola Co.&0.2343&0.0223&0.1686&0.2719&&0.2576&0.0163&0.1686&0.2731&&0.2579&0.0123&0.1686&0.2731  \\
Minnesota Mining \& MFG Co.&0.2844&0.0196&0.2399&0.3407&&0.2873&0.0099&0.2399&0.3407&&0.2802&0.0117&0.2399&0.3407  \\
Philip Morris Cos Inc.&0.1369&0.0168&0.0983&0.1673&&0.1700&0.0206&0.0983&0.1919&&0.1729&0.0155&0.0983&0.1919 \\
Pepsico Inc.&0.1634&0.0132&0.1483&0.2106&&0.1535&0.0083&0.1448&0.2106&&0.1512&0.0067&0.1434&0.2106 \\
Procter \& Gamble Co.&0.2284&0.0292&0.1434&0.2673&&0.2461&0.0169&0.1434&0.2673&&0.2413&0.0141&0.1434&0.2673  \\
Pharmacia Corp.&0.1279&0.0104&0.0863&0.1432&&0.1588&0.0192&0.0863&0.1822&&0.1643&0.0149&0.0863&0.1822 \\
Schering-Plough Corp.&0.2195&0.0190&0.1920&0.2863&&0.2179&0.0103&0.1920&0.2863&&0.2107&0.0123&0.1877&0.2863 \\
Texaco Inc.&0.4355&0.0195&0.3389&0.4906&&0.4500&0.0142&0.3389&0.4906&&0.4515&0.011&0.3389&0.4906 \\
Texas Instruments Inc.&0.0327&0.0027&0.0243&0.0369&&0.0369&0.0033&0.0243&0.0414&&0.0371&0.0027&0.0243&0.0414  \\
United Technologies Corp&0.1570&0.0153&0.1298&0.2182&&0.1562&0.0075&0.1298&0.2182&&0.1511&0.0084&0.1298&0.2182 \\
Walgreen Co.&0.0937&0.0112&0.0808&0.1384&&0.0837&0.0071&0.0776&0.1384&&0.0786&0.0078&0.0669&0.1384  \\
\hline

\end{tabular}

\caption{\label{table:lambda_detail} This table gives the average ({\it mean}),
the standard deviation ({\it std.}), the minimum ({\it min.}) and the maximum
({\it max.}) values of the coefficient of lower tail dependence estimated over
the first centile, quintile and decile during the entire time interval from
July 1962 to December 2000, under the assumption that the tail index equals
three.}

\end{center}
\end{table}

\end{landscape}

\begin{table}
\begin{center}

{\bf Coefficients of tail dependence during the time interval from July
1962 to December 1979}
\vspace{1cm}

\begin{tabular}{lcccrccc}
\hline
 & \multicolumn{3}{c}{Negative Tail} &  & \multicolumn{3}{c}{Positive Tail} \\
\cline{2-4}
\cline{6-8}
 & $\alpha=3$ & $\alpha=3.5$ & $\alpha=4$ &  & $\alpha=3$ & $\alpha=3.5$ & $\alpha=4$ \\
\hline
Abbott Labs & 0.12 & 0.09 & 0.06 &  & 0.11 & 0.08 & 0.06 \\
American Home Products Corp. & 0.22 & 0.18 & 0.15 &  & 0.25 & 0.22 & 0.19 \\
Boeing Co. & 0.16 & 0.13 & 0.10 &  & 0.13 & 0.10 & 0.07 \\
Bristol-Myers Squibb Co. & 0.22 & 0.19 & 0.16 &  & 0.28 & 0.25 & 0.23 \\
Chevron Corp. & 0.21 & 0.17 & 0.14 &  & 0.26 & 0.23 & 0.20 \\
Du Pont (E.I.) de Nemours \& Co. & 0.38 & 0.37 & 0.35 &  & 0.37 & 0.35 & 0.33 \\
Disney (Walt) Co. & 0.24 & 0.20 & 0.17 &  & 0.23 & 0.19 & 0.16 \\
General Motors Corp. & 0.39 & 0.37 & 0.35 &  & 0.48 & 0.47 & 0.47 \\
Hewlett-Packard Co. & 0.15 & 0.12 & 0.09 &  & 0.23 & 0.20 & 0.17 \\
Coca-Cola Co. & 0.26 & 0.22 & 0.19 &  & 0.26 & 0.23 & 0.20 \\
Minnesota Mining \& MFG Co. & 0.35 & 0.32 & 0.30 &  & 0.35 & 0.33 & 0.31 \\
Philip Morris Cos Inc. & 0.25 & 0.22 & 0.19 &  & 0.20 & 0.17 & 0.14 \\
Pepsico Inc. & 0.15 & 0.12 & 0.09 &  & 0.17 & 0.14 & 0.11 \\
Procter \& Gamble Co. & 0.23 & 0.19 & 0.16 &  & 0.24 & 0.21 & 0.18 \\
Pharmacia Corp. & 0.23 & 0.19 & 0.16 &  & 0.26 & 0.23 & 0.20 \\
Schering-Plough Corp. & 0.21 & 0.18 & 0.15 &  & 0.20 & 0.17 & 0.14 \\
Texaco Inc. & 0.47 & 0.46 & 0.46 &  & 0.49 & 0.49 & 0.49 \\
Texas Instruments Inc. & 0.06 & 0.04 & 0.03 &  & 0.07 & 0.05 & 0.03 \\
United Technologies Corp & 0.13 & 0.10 & 0.07 &  & 0.13 & 0.10 & 0.07 \\
Walgreen Co. & 0.03 & 0.02 & 0.01 &  & 0.02 & 0.01 & 0.01 \\
\hline
\end{tabular}

\caption{\label{table:lambda_62_79} This table summarizes the mean values
over the first centile of the distribution of the coefficients of (upper or
lower) tail dependence for the positive and negative tails during  the time
interval from July 1962 to December 1979, for three values of the tail
index $\alpha =$ 3, 3.5, 4.}

\end{center}
\end{table}

\begin{table}
\begin{center}

{\bf Coefficients of tail dependence during the time interval from January
1980 to December 2000}
\vspace{1cm}

\begin{tabular}{lcccrccc}
\hline
 & \multicolumn{3}{c}{Negative Tail} &  & \multicolumn{3}{c}{Positive Tail} \\
\cline{2-4}
\cline{6-8}
 & $\alpha=3$ & $\alpha=3.5$ & $\alpha=4$ &  & $\alpha=3$ & $\alpha=3.5$ & $\alpha=4$ \\
\hline
Abbott Labs & 0.20 & 0.17 & 0.14 &  & 0.16 & 0.13 & 0.10 \\
American Home Products Corp. & 0.12 & 0.09 & 0.06 &  & 0.10 & 0.08 & 0.05 \\
Boeing Co. & 0.14 & 0.11 & 0.08 &  & 0.10 & 0.07 & 0.05 \\
Bristol-Myers Squibb Co. & 0.32 & 0.29 & 0.26 &  & 0.25 & 0.21 & 0.19 \\
Chevron Corp. & 0.18 & 0.14 & 0.11 &  & 0.13 & 0.09 & 0.07 \\
Du Pont (E.I.) de Nemours \& Co. & 0.23 & 0.20 & 0.17 &  & 0.16 & 0.13 & 0.10 \\
Disney (Walt) Co. & 0.16 & 0.13 & 0.10 &  & 0.15 & 0.12 & 0.09 \\
General Motors Corp. & 0.26 & 0.22 & 0.19 &  & 0.20 & 0.16 & 0.13 \\
Hewlett-Packard Co. & 0.19 & 0.15 & 0.13 &  & 0.21 & 0.18 & 0.15 \\
Coca-Cola Co. & 0.24 & 0.20 & 0.18 &  & 0.20 & 0.17 & 0.14 \\
Minnesota Mining \& MFG Co. & 0.26 & 0.23 & 0.20 &  & 0.20 & 0.17 & 0.14 \\
Philip Morris Cos Inc. & 0.11 & 0.08 & 0.06 &  & 0.11 & 0.08 & 0.06 \\
Pepsico Inc. & 0.17 & 0.14 & 0.11 &  & 0.14 & 0.11 & 0.09 \\
Procter \& Gamble Co. & 0.24 & 0.21 & 0.18 &  & 0.20 & 0.16 & 0.13 \\
Pharmacia Corp. & 0.10 & 0.08 & 0.05 &  & 0.10 & 0.07 & 0.05 \\
Schering-Plough Corp. & 0.23 & 0.20 & 0.17 &  & 0.16 & 0.13 & 0.10 \\
Texaco Inc. & 0.43 & 0.42 & 0.41 &  & 0.31 & 0.28 & 0.26 \\
Texas Instruments Inc. & 0.02 & 0.01 & 0.01 &  & 0.02 & 0.01 & 0.01 \\
United Technologies Corp & 0.20 & 0.16 & 0.14 &  & 0.18 & 0.14 & 0.11 \\
Walgreen Co. & 0.15 & 0.12 & 0.09 &  & 0.09 & 0.07 & 0.05 \\
\hline
\end{tabular}

\caption{\label{table:lambda_80_00} This table summarizes the mean values
over the first centile of the distribution of the coefficients of (upper or
lower) tail dependence for the positive and negative tails during  the time
interval from January 1980 to December 2000, for three values of the tail
index $\alpha =$ 3, 3.5, 4.}

\end{center}
\end{table}

\begin{table}
\begin{center}

{\bf Coefficients of tail dependence during the time interval from July
1962 to December 2000}
\vspace{1cm}

\begin{tabular}{lcccrccc}
\hline
 & \multicolumn{3}{c}{Negative Tail} &  & \multicolumn{3}{c}{Positive Tail} \\
\cline{2-4}
\cline{6-8}
 & $\alpha=3$ & $\alpha=3.5$ & $\alpha=4$ &  & $\alpha=3$ & $\alpha=3.5$ & $\alpha=4$ \\
\hline
Abbott Labs & 0.17 & 0.13 & 0.11 &  & 0.15 & 0.12 & 0.09 \\
American Home Products Corp. & 0.14 & 0.11 & 0.08 &  & 0.15 & 0.11 & 0.09 \\
Boeing Co. & 0.14 & 0.10 & 0.08 &  & 0.10 & 0.07 & 0.05 \\
Bristol-Myers Squibb Co. & 0.27 & 0.24 & 0.21 &  & 0.27 & 0.24 & 0.21 \\
Chevron Corp. & 0.19 & 0.15 & 0.12 &  & 0.17 & 0.13 & 0.10 \\
Du Pont (E.I.) de Nemours \& Co. & 0.25 & 0.22 & 0.19 &  & 0.23 & 0.19 & 0.16 \\
Disney (Walt) Co. & 0.18 & 0.14 & 0.11 &  & 0.17 & 0.13 & 0.11 \\
General Motors Corp. & 0.26 & 0.23 & 0.20 &  & 0.24 & 0.21 & 0.18 \\
Hewlett-Packard Co. & 0.17 & 0.14 & 0.11 &  & 0.23 & 0.19 & 0.16 \\
Coca-Cola Co. & 0.23 & 0.20 & 0.17 &  & 0.23 & 0.20 & 0.17 \\
Minnesota Mining \& MFG Co. & 0.28 & 0.25 & 0.23 &  & 0.25 & 0.22 & 0.19 \\
Philip Morris Cos Inc. & 0.14 & 0.10 & 0.08 &  & 0.14 & 0.11 & 0.08 \\
Pepsico Inc. & 0.16 & 0.13 & 0.10 &  & 0.16 & 0.12 & 0.10 \\
Procter \& Gamble Co. & 0.23 & 0.20 & 0.17 &  & 0.22 & 0.18 & 0.15 \\
Pharmacia Corp. & 0.13 & 0.10 & 0.07 &  & 0.14 & 0.10 & 0.08 \\
Schering-Plough Corp. & 0.22 & 0.19 & 0.16 &  & 0.19 & 0.15 & 0.12 \\
Texaco Inc. & 0.44 & 0.42 & 0.41 &  & 0.37 & 0.35 & 0.33 \\
Texas Instruments Inc. & 0.03 & 0.02 & 0.01 &  & 0.03 & 0.02 & 0.01 \\
United Technologies Corp & 0.16 & 0.12 & 0.10 &  & 0.15 & 0.12 & 0.09 \\
Walgreen Co. & 0.09 & 0.07 & 0.05 &  & 0.06 & 0.04 & 0.03 \\
\hline
\end{tabular}

\caption{\label{table:lambda_62_00} This table summarizes the mean values
over the first centile of the distribution of the coefficients of (upper or
lower) tail dependence for the positive and negative tails during  the time
interval from July 1962 to December 2000, for three values of the tail
index $\alpha =$ 3, 3.5, 4.}

\end{center}
\end{table}

\begin{table}
\begin{center}

{\bf Comparison of the estimated coefficient of lower tail dependence with the
realized extreme losses}
\vspace{1cm}

\begin{tabular}{lccccccc}
\hline
 & \multicolumn{3}{c}{July 1962 - Dec. 1979} &  & \multicolumn{3}{c}{Jan.1980 - Dec. 2000} \\
\cline{2-4}
\cline{6-8}
 & Extremes & $\lambda_-$ & p-value &  & Extremes & $\lambda_-$ & p-value \\
\hline
Abbott Labs & 0 & 0.12 & 0.2937 &  & 4 & 0.20 & 0.0904 \\
American Home Products Corp. & 1 & 0.22 & 0.2432 &  & 2 & 0.12 & 0.2247 \\
Boeing Co. & 0 & 0.16 & 0.1667 &  & 3 & 0.14 & 0.1176 \\
Bristol-Myers Squibb Co. & 2 & 0.22 & 0.2987 &  & 4 & 0.32 & 0.2144 \\
Chevron Corp. & 3 & 0.21 & 0.2112 &  & 4 & 0.18 & 0.0644 \\
Du Pont (E.I.) de Nemours \& Co. & 0 & 0.38 & 0.0078 &  & 4 & 0.23 & 0.1224 \\
Disney (Walt) Co. & 2 & 0.24 & 0.2901 &  & 2 & 0.16 & 0.2873 \\
General Motors Corp. & 2 & 0.39 & 0.1345 &  & 4 & 0.26 & 0.1522 \\
Hewlett-Packard Co. & 0 & 0.15 & 0.1909 &  & 2 & 0.19 & 0.3007 \\
Coca-Cola Co. & 2 & 0.26 & 0.2765 &  & 5 & 0.24 & 0.0494 \\
Minnesota Mining \& MFG Co. & 2 & 0.35 & 0.1784 &  & 4 & 0.26 & 0.1571 \\
Philip Morris Cos Inc. & 1 & 0.25 & 0.1841 &  & 2 & 0.11 & 0.2142 \\
Pepsico Inc. & 2 & 0.15 & 0.2795 &  & 5 & 0.17 & 0.0141 \\
Procter \& Famble Co. & 1 & 0.23 & 0.2245 &  & 3 & 0.24 & 0.2447 \\
Pharmacia Corp. & 2 & 0.23 & 0.2956 &  & 4 & 0.10 & 0.0128 \\
Schering-Plough Corp. & 0 & 0.21 & 0.0946 &  & 4 & 0.23 & 0.1224 \\
Texaco Inc. & 1 & 0.47 & 0.0161 &  & 3 & 0.43 & 0.1862 \\
Texas Instruments Inc. & 0 & 0.06 & 0.5222 &  & 2 & 0.02 & 0.0212 \\
United Technologies Corp & 1 & 0.13 & 0.3728 &  & 4 & 0.20 & 0.0870 \\
Walgreen Co. & 1 & 0.03 & 0.2303 &  & 3 & 0.15 & 0.1373 \\
\hline
\end{tabular}

\caption{\label{table:extremes} This table gives, for the time intervals from
July 1962 to December 1979 and from January 1980 to December 2000, the number
of losses within the ten largest losses incured by an asset which have occured
together with one of the ten largest losses of the Standard \& Poor's 500
index during the same time interval. The probabilty of occurence of such a
realisation is given by the p-value derived from the binomial law
(\ref{eq:binomial}) with parameter $\lambda_-$.}

\end{center}
\end{table}

\begin{table}
\begin{center}

{\bf Comparison of the estimated coefficient of lower tail dependence with the
realized non-comonotonic extreme losses}
\vspace{1cm}

\begin{tabular}{lccccccc}
\hline
 & \multicolumn{3}{c}{July 1962 - Dec. 1979} &  & \multicolumn{3}{c}{Jan.1980 - Dec. 2000} \\
\cline{2-4}
\cline{6-8}
 & Extremes & $\lambda_-$ & p-value &  & Extremes & $\lambda_-$ & p-value \\
\hline
Abbott Labs & 0 & 0.12 & 0.2937 &  & 4 & 0.20 & 0.0904 \\
American Home Products Corp. & 1 & 0.22 & 0.2432 &  & 1 & 0.12 & 0.3828 \\
Boeing Co. & 0 & 0.16 & 0.1667 &  & 3 & 0.14 & 0.1176 \\
Bristol-Myers Squibb Co. & 2 & 0.22 & 0.2987 &  & 3 & 0.32 & 0.2653 \\
Chevron Corp. & 3 & 0.21 & 0.2112 &  & 3 & 0.18 & 0.1708 \\
Du Pont (E.I.) de Nemours \& Co. & 0 & 0.38 & 0.0078 &  & 3 & 0.23 & 0.2342 \\
Disney (Walt) Co. & 2 & 0.24 & 0.2901 &  & 1 & 0.16 & 0.3300 \\
General Motors Corp. & 2 & 0.39 & 0.1345 &  & 3 & 0.26 & 0.2536 \\
Hewlett-Packard Co. & 0 & 0.15 & 0.1909 &  & 1 & 0.19 & 0.2880 \\
Coca-Cola Co. & 1 & 0.26 & 0.1782 &  & 4 & 0.24 & 0.1318 \\
Minnesota Mining \& MFG Co. & 2 & 0.35 & 0.1784 &  & 3 & 0.26 & 0.2561 \\
Philip Morris Cos Inc. & 1 & 0.25 & 0.1841 &  & 2 & 0.11 & 0.2142 \\
Pepsico Inc. & 2 & 0.15 & 0.2795 &  & 5 & 0.17 & 0.0141 \\
Procter \& Famble Co. & 1 & 0.23 & 0.2245 &  & 3 & 0.24 & 0.2447 \\
Pharmacia Corp. & 2 & 0.23 & 0.2956 &  & 4 & 0.10 & 0.0128 \\
Schering-Plough Corp. & 0 & 0.21 & 0.0946 &  & 3 & 0.23 & 0.2342 \\
Texaco Inc. & 1 & 0.47 & 0.0161 &  & 3 & 0.43 & 0.1862 \\
Texas Instruments Inc. & 0 & 0.06 & 0.5222 &  & 1 & 0.02 & 0.1922 \\
United Technologies Corp & 1 & 0.13 & 0.3728 &  & 3 & 0.20 & 0.2001 \\
Walgreen Co. & 1 & 0.03 & 0.2303 &  & 3 & 0.15 & 0.1373 \\
\hline
\end{tabular}

\caption{\label{table:comonotonic} This table gives, for the time intervals
from July 1962 to December 1979 and from January 1980 to December 2000, the
number of losses within the ten largest losses incured by an asset which have
occured together with one of the ten largest losses of the Standard \& Poor's
500 index during the same time interval, provided that the losses are not both
the largest of each series. The probabilty of occurence of such a realisation
is given by the p-value derived from the binomial law (\ref{eq:binomial}) with
parameter $\lambda_-$.}

\end{center}
\end{table}

\end{document}